# ENVIRONMENTALLY FRIENDLY RENORMALIZATION

# IN FINITE-TEMPERATURE FIELD THEORY[†]

M.A. van Eijck[1], Denjoe O'Connor[2] and C.R.Stephens[2]
[1]Institute for Theoretical Physics, University of Amsterdam
Valckenierstraat 65, 1018 XE Amsterdam, The Netherlands
[2]DIAS, 10 Burlington Road, Dublin 4, Ireland

## INTRODUCTION

Phase transitions in relativistic quantum field theory are a topic of great theoretical and practical interest. Some of the more interesting ones are the deconfinement phase transition in QCD and the electroweak phase transition, the latter having generated much attention recently due to its possible role in explaining the baryon asymmetry of the universe.

It is well known that a quantitative description of finite temperature phase transitions presents grave difficulties, not least because of the infrared divergences (IR) that have been known since the work of Dolan and Jackiw and Weinberg[1]. It is first important to understand the physical origin of such difficulties in order that they may be overcome. In past work[2,3] we have emphasized that the origin of such difficulties lies in the fact that the effective degrees of freedom (EDOF) of a system can change qualitatively as a function of temperature. For instance, above the QCD deconfinement phase transition quarks are free whereas below it they exist only as bound states in mesons and hadrons. As a function of temperature then there is a drastic change in the nature of the EDOF from quarks to baryons. The IR problems intrinsic to a description of the phase transition in $\lambda\phi^4$ theory have an analogous origin. A perturbation expansion in terms of the zero temperature coupling in this case breaks down near the critical temperature because the EDOF there are three dimensional rather than four dimensional. A popular approach to this problem has been to *assume* dimensional reduction and proceed to describe the critical region as an effective three dimensional theory using $\varepsilon$ expansion methods for instance[5,6]. However, whether a particular field theory will dimensionally reduce or not is something one should derive not assume. In other words the ideal would be to develop a quantitatively reliable method based solely on information about the zero temperature theory.

---

[†] Talk presented by CRS at NATO workshop on the Electroweak Phase Transition and the Early Universe, Sintra, March 1994. To be published in the Proceedings by Plenum Press.

We have shown previously[2,3,4] that environmentally friendly renormalization provides a very powerful tool for the qualitative and quantitative description of finite temperature phase transitions. The general gist of the approach is the following: the EDOF (fluctuations) in a system are very often sensitive to the "environment" in which they live. Temperature is just one example of an environmental variable which drastically affects the EDOF. Given that one is interested in how the physical behaviour of systems change as a function of "scale" (momentum, mass, temperature, magnetic field etc.), and that the renormalization group (RG) is an extremely powerful tool for investigating physics as a function of scale, then one should implement a RG which is environmentally friendly in the sense that viewed as a coarse graining procedure it coarse grains as closely as possible the true environment dependent EDOF. This means that the RG should depend on any relevant environmental parameters in the problem. In particular in the context of finite temperature field theory it should be explicitly temperature dependent.

Although one can fruitfully ground one's intuition in a coarse graining point of view it is important to point out that the environmentally friendly RG's we employ are not in practice of the Wilsonian type such as the "average effective action"[7]. Rather, they exploit the reparametrization invariance of field theories, in that one may parameterize the results of one's experiments in terms of different sets of renormalized parameters. Although non-perturbatively speaking physical results cannot depend on the particular renormalization scheme chosen, when one resorts to an approximation scheme such as perturbation theory one finds that the reliability of the approximation scheme may be quite crucially dependent on the parameters chosen. More specifically one should always try to implement an environmentally friendly reparametrization. We believe that an approach based on a reparametrization point of view, although less intuitive, is more powerful than that based on coarse graining. The latter is always concerned with integrating out degrees of freedom as a function of an effective cutoff. The RG flow is always with respect to this cutoff. In contrast, as has been emphasized previously[3], and as will also be seen herein, the reparametrization point of view, in principle, allows for the RG flow to be with respect to any parameter in the problem. One can even implement several RG's at the same time as we will see below, where RG's based on an arbitrary fiducial momentum scale $\kappa$ and on an arbitrary fiducial temperature scale $\tau$ are used together. An important feature of environmentally friendly RG's is that the fixed points of such RG's will coincide with all the points of conformal invariance of a particular theory.

# ENVIRONMENTALLY FRIENDLY RENORMALIZATION

We will briefly review in this section environmentally friendly renormalization in the context of finite temperature field theory; more specifically we will consider two environmentally friendly RG's. In the first an RG is generated within which an arbitrary, fiducial mass is the running parameter and in the second an arbitrary, fiducial temperature. We will refer to the latter as "running the environment". As discussed in the introduction, in our formalism a particular RG is associated with reparametrizations of the "couplings" (masses, interaction strengths etc.) of a field theory. Here we exhibit a reparametrization from bare to renormalized parameters that yields perturbatively valid answers for any value of the temperature, and in particular in the critical regime $T \sim T_c$.

For the most part we will consider $\lambda \phi^4$ theory in this paper, defined by the Euclidean action

$$S[\phi_B] = \int_0^{\frac{1}{T}} \int d^4x \left( \frac{1}{2}(\nabla \phi_B)^2 + \frac{1}{2}m_B^2 \phi_B^2 + \frac{\lambda_B}{4!}\phi_B^4 + j_B(x)\phi_B + \frac{1}{2}t_B(x)\phi_B^2 \right)$$

which proffers a paradigm for considerations of the Higgs sector of the standard model, though we will consider gauge theories briefly in a later section. $j_B(x)$ and $t_B(x)$ are "sources" which generate correlators of $\phi_B$ and $\phi_B^2$ respectively. Note that these sources could represent real effects. Here a constant $t_B$ will be a mass parameter. We will work initially in the symmetric phase ($j_B = 0$, $T > T_c$) and consider reparametrizations obtained from a set of normalization conditions. For the first case, running a fiducial mass, we choose the following

$$\Gamma^{(2)}(p=0, m(T) = m_{min}, \lambda, T, \kappa) = m_{min}^2 \tag{1}$$

$$\frac{\partial \Gamma^{(2)}}{\partial \vec{p}^2}(p_0 = 0, \vec{p}, m(T) = \kappa, \lambda, T, \kappa)|_{\vec{p}=0} = 1 \tag{2}$$

$$\Gamma^{(4)}(p=0, m(T) = \kappa, \lambda, T, \kappa) = \lambda \tag{3}$$

$$\frac{\partial \Gamma^{(2)}}{\partial t}(p=0, m(T) = \kappa, \lambda, T, \kappa) = 1 \tag{4}$$

where $t$ is the renormalized mass parameter. Note that we are here parameterizing the system in terms of an an arbitrary, fiducial screening length, $\kappa^{-1}$, not the actual screening length, $m^{-1}$, which would lead to a Callan-Symanzik type equation. In Eq. (1) $m_{min}$ represents the value of the physical mass when $t=0$ and is an RG invariant point. If $m_{min} = 0$ the theory exhibits a second order phase transition. Eq.'s (2-4) fix the multiplicative renormalization constants $Z_\phi$, $Z_{\phi^2}$ and $Z_\lambda$ which renormalize $\phi$, the composite operator $\phi^2$ and $\lambda$ respectively. Using these conditions, and implementing a [2,1] Padé resummation of the two loop Wilson functions one finds[3]

$$\beta(h, \frac{T}{\kappa}) = -\varepsilon(\frac{T}{\kappa})h + \frac{h^2}{1 + 4\left(\frac{(5N+22)}{(N+8)^2}f_1(\frac{T}{\kappa}) - \frac{(N+2)}{(N+8)^2}f_2(\frac{T}{\kappa})\right)h} \tag{5}$$

$$\gamma_{\phi^2}(h, \frac{T}{\kappa}) = \frac{(N+2)}{(N+8)} \frac{h}{1 + \frac{6}{(N+8)}\left(f_1(\frac{T}{\kappa}) - \frac{1}{3}f_2(\frac{T}{\kappa})\right)h} \tag{6}$$

and
$$\gamma_\phi(h, \frac{T}{\kappa}) = 2\frac{(N+2)}{(N+8)^2} f_2(\frac{T}{\kappa}) h^2 \tag{7}$$

where $\varepsilon(\frac{T}{\kappa}) = 1 + \kappa \frac{d}{d\kappa} \ln(\sum_n m^{-3})$

$$f_1(\frac{T}{\kappa}) = 2\frac{\sum_{n_1,n_2}(\frac{1}{m_1^3}(\frac{1}{M} - \frac{1}{2m_2}) + \frac{1}{m_1 M^2}(\frac{1}{m_1} + \frac{2}{m_2}))}{(\sum_n \frac{1}{m^3})^2} \quad \text{and} \quad f_2(\frac{T}{\kappa}) = 4\frac{\sum_{n_1,n_2} \frac{1}{M^3 m_1}}{(\sum_n \frac{1}{m^3})^2}$$

with

$$m_i = (1 + \frac{4\pi^2 n_i^2 \kappa^2}{T^2})^{\frac{1}{2}}, \quad m_{12} = (1 + \frac{4\pi^2 \kappa^2}{T^2}(n_1 + n_2)^2)^{\frac{1}{2}}, \quad M = m_1 + m_2 + m_{12}$$

$\gamma_\phi$ and $\gamma_{\phi^2}$ are the anomalous dimensions of $\phi$ and $\phi^2$ respectively. In Eq.'s (5-7) the coupling $h$, or floating coupling[2,3], is defined via the relation $h = a_2(\frac{T}{\kappa})\lambda(\kappa)$, where $a_2$ is the coefficient of $\lambda^2$ in $\beta(\lambda)$. After the characteristic equations are solved we can use our freedom in choosing the arbitrary fiducial mass $\kappa$ to be the physical finite temperature mass $m(T)$. An important feature of Eq. (5) is that it exhibits more than one fixed point. As $\frac{T}{m(T)} \to 0$, one obtains the Gaussian fixed point as expected in four dimensions. As $T \to T_c$, i.e. $m(T) \to 0$ one finds a non-trivial fixed point, $h = 1.732$ for $N = 1$, for instance. The value of the fixed point and the corresponding critical exponents are in exact agreement with corresponding two loop Padé resummed results[8] in three dimensional critical phenomena. Observe that the coupling $\lambda \to 0$ as the critical temperature is approached. As emphasized in O' Connor and Stephens[2,3] (and later also by Tetradis and Wetterich[7]) this however does not mean that the theory is non-interacting in this limit but only that $\lambda$ is an inappropriate measure of the effective interactions, the floating coupling $h$ being a more transparent measure.

Notice that we have derived the fact that the theory is three dimensional near the phase transition instead of assuming it, as we are able to examine the complete crossover between four and three dimensional behaviour, as $\frac{T}{m(T)}$ varies between 0 and $\infty$, in a perturbatively controllable fashion. Here we are probing different values of $m(T)$ by varying the mass parameter $t$ for fixed temperature, below we will use a complementary approach where $T$ is varied for fixed $t$.

One can think of $d_{eff} = 4 - \varepsilon(\frac{T}{m(T)})$, which interpolates between four and three when $\frac{T}{m(T)}$ varies between 0 and $\infty$, as a measure of the effective dimension of the system. Near $T = 0$ the EDOF of the problem are four dimensional and near $T = T_c$, three dimensional. The power of our approach is that we have implemented a reparametrization which is temperature dependent in such a fashion that it tracks the evolving nature of the EDOF between the four and three dimensional limits. In the above we considered only the symmetric phase. An environmentally friendly $\kappa$ RG can also be implemented for the broken phase[9].

We see then how environmentally friendly methods can yield a perturbatively reliable description of the finite temperature behaviour from $T = 0$ to $T = \infty$. The description that we have obtained though is in terms of the finite temperature screening length and the finite temperature coupling rather than the parameters of the zero temperature theory, which are, after all, the ones we would have experimental access to. Additionally there are certain quantities, such as the critical temperature itself, which are very difficult to obtain from this RG. We consider therefore a complementary approach wherein we renormalize at a fiducial temperature $\tau$. Such a RG was first introduced by Matsumoto et al.[10]. Here we will show how such a RG can be utilized and exploited in an environmentally friendly context. Once again the reparametrization we will use is defined by a set of normalization conditions which we take to be

$$\Gamma^{(2)}(p=0, M, \lambda, \bar{\phi}, T=\tau) = M^2 \tag{8}$$

$$\Gamma^{(4)}(p=0, M, \lambda, \bar{\phi}, T=\tau) = \lambda \tag{9}$$

$$\frac{d}{d\vec{p}^2}\Gamma(p_0=0, \vec{p}, M, \lambda, \bar{\phi}, T=\tau)\bigg|_{\vec{p}=0} = 1 \tag{10}$$

where $M^{-1}$ is the finite temperature screening length at the fiducial temperature $\tau$ and $\bar{\phi}$ represents the minimum of the effective potential at that temperature, i.e. it satisfies the equation of state

$$\Gamma^{(1)}(p=0, M, \lambda, \bar{\phi}, T=\tau) = 0. \tag{11}$$

The beta functions, obtained by differentiating these renormalization conditions with respect to $\tau$ for fixed bare parameters, depend on derivatives $d\bar{\phi}/d\tau$. These can be eliminated using the equation of state. The differential equations describing an infinitesimal change in the normalization point are

$$\tau\frac{dM^2}{d\tau} = \beta_M(\lambda, M, \tau) \qquad \tau\frac{d\lambda}{d\tau} = \beta_\lambda(\lambda, M, \tau) \tag{12}$$

The functions $\beta_M$ and $\beta_\lambda$ take different functional forms above and below the critical temperature. To one loop for $T > T_c$

$$\beta_M = \frac{\lambda}{2}\tau\frac{\partial \bigcirc}{\partial \tau} \qquad \beta_\lambda = -\frac{3}{2}\lambda^2\tau\frac{d\bigcirc}{d\tau} \tag{13}$$

and below $T_c$ we find

$$\begin{aligned}\beta_M &= -\lambda\left(\tau\frac{\partial \bigcirc}{\partial \tau} + \frac{3}{2}M^2\tau\frac{\partial \bigcirc}{\partial \tau} + \frac{1}{2}M^4\tau\frac{\partial \bigcirc}{\partial \tau}\right) \\ \beta_\lambda &= -\frac{3}{2}\lambda^2\tau\frac{d}{d\tau}\left(\bigcirc - \frac{34}{3}M^2\bigcirc + 18M^4\bigcirc\right)\end{aligned} \tag{14}$$

We have presented our results in diagramatic form as this renders the structure of these expressions more readily apparent and easily adoptable to other situations. Our diagrammatic notation is that the dots on a diagram represent the location of insertions in that diagram. The insertions could arise from different sources. All the one loop diagrams can be obtained from the basic diagram $\bigcirc$ which using dimensional regularization is given by

$$\bigcirc = -\frac{\Gamma(-\frac{d}{2})M^d}{(4\pi)^{d/2}} - \frac{2\tau^d}{(4\pi)^{(d-1)/2}\Gamma(\frac{d+1}{2})} \int_0^\infty \frac{dq}{\sqrt{q^2 + \frac{M^2}{\tau^2}}} \frac{1}{e^{\sqrt{q^2+\frac{M^2}{\tau^2}}} - 1} \tag{15}$$

The derivative with respect to $M^2$ of a diagram with $k$ dots gives $-k$ times a diagram with $k+1$ dots. Note that $M$ is retained in a non-perturbative manner while $\lambda$ is retained only to lowest non-trivial order. This accounts for the asymmetry between the partial derivatives in $\beta_M$ but total derivatives in $\beta_\lambda$. Such a procedure, which can be consistently carried out to higher order, automatically uses the fully dressed mass, $M$, and in particular sums all daisy insertions on all the internal lines of the $\beta$ function diagrams. Such an artifice is crucial in that without it the tadpole insertions, or "shift", which constitute the largest temperature effects would ensure a breakdown in perturbation theory. In using such a procedure we differ substantially from other authors[11] who all neglect the term $\tau \frac{dM^2}{d\tau}$ in $\beta_\lambda$ in Eq.'s (17) and (18). As demonstrated below this is inconsistent in the vicinity of the critical point.

We will now examine the structure of our flow equations near the critical point. The asymptotic expansion for small $M$ of (15) is

$$\bigcirc = \tau^d \left( -\frac{2\Gamma(d/2)\zeta(d)}{\pi^{d/2}} + \frac{\Gamma(\frac{d-2}{2})\zeta(d-2)}{2\pi^{d/2}} \frac{M^2}{\tau^2} - \frac{\Gamma(\frac{1-d}{2})}{(4\pi)^{(d-1)/2}} \left(\frac{M}{\tau}\right)^{(d-1)} \right.$$
$$\left. - \frac{\Gamma(\frac{d-4}{2})\zeta(d-4)}{(4\pi)^2 \pi^{(d-4)/2}} \left(\frac{M}{\tau}\right)^4 + \cdots \right) \tag{16}$$

for any dimension $d$ between three and four. Thus in the neighbourhood of the critical point for the four dimensional theory, we find

$$\beta_M = \lambda \left( \frac{\tau^2}{12} - \frac{M\tau}{8\pi} + \cdots \right)$$
$$\beta_\lambda = -\frac{3\tau\lambda^2}{16\pi M} \left( 1 - \frac{1}{2M^2}\tau\frac{dM^2}{d\tau} + \cdots \right) \tag{17}$$

for $T > T_c$ and

$$\beta_M = -\lambda \left( \frac{\tau^2}{6} - \frac{29M\tau}{64\pi} + \cdots \right)$$
$$\beta_\lambda = -\frac{5\lambda^2\tau}{64\pi M} \left( 1 - \frac{1}{2M^2}\tau\frac{dM^2}{d\tau} + \cdots \right) \tag{18}$$

for $T < T_c$. The differential equation for $\lambda$ is easy to solve since it is a total derivative. We find that as the critical point is approached $\lambda \to 0$ which is consistent with what we found using an RG where a fiducial mass was the running parameter. As explained above this is merely a manifestation of the fact that in the critical region a more appropriate coupling would be the effective three dimensional coupling $\frac{\lambda T}{M(T)}$. Explicitly

$$\lambda = \frac{16\pi M}{3\tau} + \cdots \quad \text{for } T > T_c \text{ and} \quad \lambda = \frac{64\pi M}{5\tau} + \cdots \quad \text{for } T < T_c \tag{19}$$

The solution of the differential equation for $\lambda$ can be substituted into that for $M$ to find, for example, for $T > T_c$

$$\frac{1}{M^2}\tau\frac{dM^2}{d\tau} = \frac{4\pi\tau}{9M} - \frac{2}{3} + \ldots \qquad (20)$$

The fact that this quantity diverges as the critical region is approached gives ample justification as to why it was necessary to use total derivatives and not partial derivatives in Eq.'s (13) and (14).

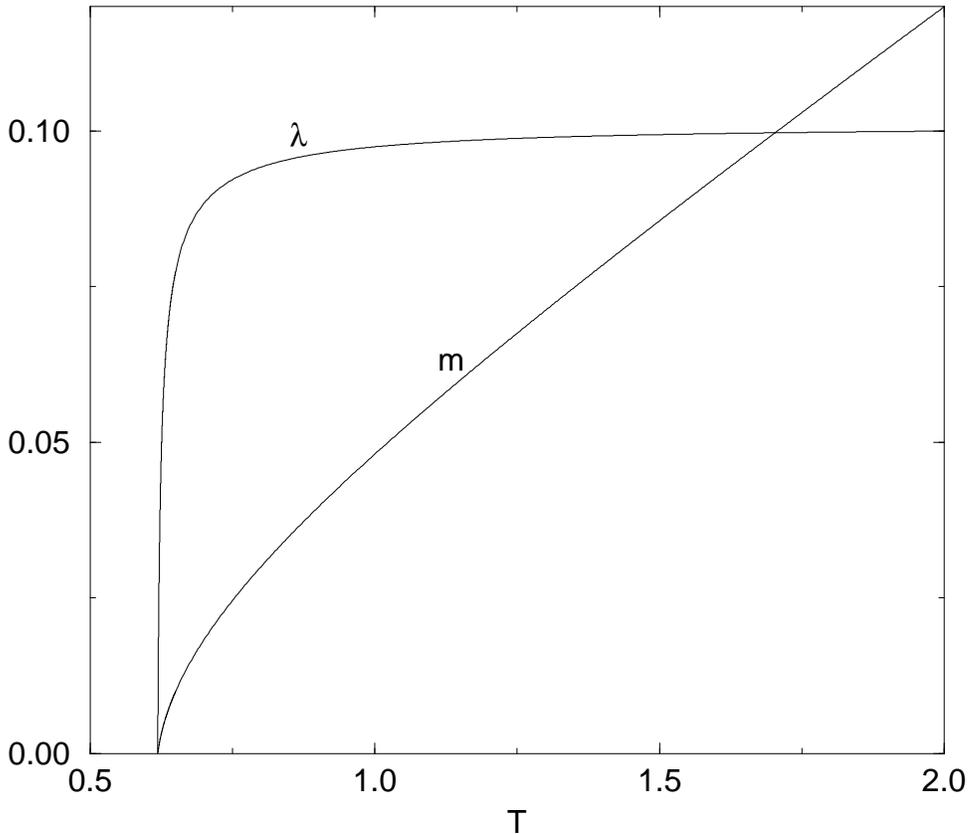

Fig. 1: Solutions of the coupled differential equations (12) for the mass $M$ and coupling $\lambda$ as a function of temperature $T$.

The solutions to the equations (12) for $T > T_c$ are plotted in Fig. 1 where we have chosen $\tau = T$, the physical temperature. As emphasized in Freire and Stephens[3] such a choice is essential if a perturbatively reliable description is to be obtained for all possible temperatures. The critical temperature is clearly visible and is a solution of $M(T_c, M(T = T_{in}), \lambda(T = T_{in})) = 0$. The critical temperature being a non-universal quantity depends on the "non-universal" initial conditions of the RG flow. We emphasize however that $T_c$ nevertheless is obtained from the RG flow. If we use Eq.'s (12) for $T < T_c$ we can choose the initial temperature $T_{in} = 0$ and thereby relate $T_c$ to the zero temperature parameters. As $M$ goes continuously to zero for $T \to T_c$ we see that the phase transition is second order. The effective dimension $d_{eff}(\frac{T}{M(T)})$ provides a quantitative measure of the extent to which dimensional reduction has occurred. The latter in this setting is manifestly *derived* from the solutions of Eq. (12) for $T < T_c$ based only on a knowledge of the zero temperature parameters. Two loop results for $T \geq T_c$ have been obtained and along with a more detailed discussion of the results in the broken phase will be presented elsewhere.

Let us now make some brief comments about the effective potential. The effective potential is convex, as it must be on physical grounds, the region between the minima being physically inaccessible for any constant background field. The minima of the effective potential are obtained from the equation of state (11). We find to one loop that

$$\bar{\phi} = \pm \sqrt{\frac{3}{\lambda(T)}} M(T) \left[ 1 + 9\lambda(T) M^2(T)(\bigcirc - \frac{3}{2}\lambda(T) M^2(T) \bigcirc) + \ldots \right] \quad (21)$$

where $M(T)$ and $\lambda(T)$, solutions of equations (12), are functions of the zero temperature parameters and $T$.

The two environmentally friendly RG's we have shown here are complementary. Although they both in principle give a complete description of the finite temperature behaviour some questions are more easily addressed using one versus the other, in particular the $\tau$ RG allows for a completely RG determination of the critical temperature. Together the two RG's offer perturbatively reliable answers for basically any physical quantity.

## NON-ABELIAN GAUGE FIELDS

We now discuss briefly some aspects of finite temperature non-abelian gauge fields, further details may be found in other work[4]. Here we will concentrate on the gauge coupling in the magnetic sector. We use as a renormalization condition that the static (i.e. zero energy), spatial three-gluon vertex equals the tree-level vertex in the symmetric momentum configuration

$$\Gamma^{abc}_{ijk}(p^0_i = 0, \vec{p}_i, g_{\kappa,\tau}, T = \tau)\big|_{\substack{\text{symm.}\\ \kappa}} = g_{\kappa,\tau} f^{abc} \left[ \eta_{ij}(p_1 - p_2)_k + cycl. \right]. \quad (22)$$

where $\eta_{ij}$ is the metric and $f_{abc}$ are the structure constants of the gauge group.

In contradistiction to the previously discussed cases this chosen normalization condition depends on two flow parameters simultaneously, the momentum scale $\kappa$, and the temperature scale $\tau$. Therefore we can perform a RG analysis with respect to both parameters, i.e. we can run more than one environmental parameter at the same time. The $\tau$ RG is needed to draw conclusions about the temperature dependence of the coupling. This cannot be done using the $\kappa$-scheme alone without assuming something about the temperature dependence of the initial value of the coupling used in solving the differential equation.

For the calculation we use the Landau gauge Background Field Feynman rules[12] resulting from the Vilkovisky-de Witt effective action in order to get rid of ambiguities arising from gauge dependence. In terms of the coupling $\alpha_{\kappa,\tau} := g^2_{\kappa,\tau}/4\pi^2$ the $\beta$ function equations are

$$\kappa \frac{d\alpha_{\kappa,\tau}}{d\kappa}\bigg|_\tau = \beta_{vac} + \beta_{th}, \qquad \tau \frac{d\alpha_{\kappa,\tau}}{d\tau}\bigg|_\kappa = -\beta_{th}, \quad (23)$$

where the vacuum contribution is, as usual,

$$\beta_{vac} = \alpha_{\kappa,\tau}^2 \left( -\frac{11}{6} N_c + \frac{1}{3} N_f \right) \tag{24}$$

and where, in terms of the IR and UV convergent integrals

$$F_n^\eta = \int_0^\infty dx \frac{x^n}{e^{\kappa x/2\tau} - \eta} \left[ \log \left| \frac{x+1}{x-1} \right| - 2 \sum_{k=0}^{\frac{n}{2}-1} \frac{x^{2k+1}}{2k+1} \right]$$

and

$$G_n^\eta = \int_0^\infty dx \frac{1}{e^{\kappa x/2\tau} - \eta} \mathrm{P} \frac{x}{(x^2-1)^n}$$

the thermal contribution is given by

$$\beta_{th} = \alpha_{\kappa,\tau}^2 \left[ \left( \frac{21}{16} F_0^1 + \frac{3}{4} F_2^1 - \frac{3}{2} G_0^1 - \frac{25}{8} G_1^1 - G_2^1 \right) N_c + \left( \frac{1}{4} F_0^{-1} + \frac{3}{4} F_2^{-1} - \frac{3}{2} G_0^{-1} - G_1^{-1} \right) N_f \right]. \tag{25}$$

Because the two beta functions (23) are not exactly each other's opposite the RG improved coupling is not just a function of the ratio $\kappa/\tau$. There is another dimensionful scale (such as $\Lambda_{QCD}$) that comes from an initial condition for these differential equations. The solution of the set of coupled differential equations can be written in the form

$$\alpha_{\kappa,\tau} = \frac{1}{\left(\frac{11}{6} N_c - \frac{1}{3} N_f\right) \ln \frac{\kappa}{\Lambda_{QCD}} - f\left(\frac{\kappa}{\tau}\right)} \tag{26}$$

where the function $f$ satisfies $\beta_{th} = \alpha_{\kappa,\tau}^2 \kappa df/d\kappa$ with the initial condition $\lim_{\tau \downarrow 0} f = 0$ so that we can identify $\Lambda_{QCD}$ with the usual zero-temperature QCD scale. Actually this function $f$ can be found in terms of the functions $F$ and $G$:

$$f = \left( \frac{21}{16} F_0^1 + \frac{1}{4} F_2^1 + \frac{7}{8} G_1^1 \right) N_c + \left( \frac{1}{4} F_0^{-1} + \frac{1}{4} F_2^{-1} \right) N_f. \tag{27}$$

Fig. 2 is a contour plot of the effective coupling as a function of both momentum and temperature scale. The results are best trusted in places where the coupling is small. For physical reasons we have to restrict ourselves in any case to the region where the coupling is positive, which is below the uppermost line in the graph.

The high-temperature behaviour (i.e. for $\tau \gg \kappa$) is determined by

$$f \longrightarrow N_c \frac{21\pi^2}{16} \frac{\tau}{\kappa} + \left( \frac{11}{6} N_c - \frac{1}{3} N_f \right) \ln \frac{\kappa}{\tau} + O(1). \tag{28}$$

As emphasized in van Eijck et al.[4] for increasing temperatures at fixed momentum scale the coupling approaches an IR Landau pole, an indication that we are entering a strong-coupling regime, whereas the opposite sign would lead to asymptotic freedom in this limit. Buchmüller and Fodor[6] have recently come to a similar conclusion based on an assumed dimensional reduction in the electroweak model. Stimulated by the original belief[13] that high-temperature QCD would be asymptotically free as in the high-momentum situation, it was suggested[14] that this strong coupling in the IR would be an artifact of the one-loop calculation and that a higher-loop calculation or a resummation could change it. We however believe that this will not happen, as the sign appears quite naturally if one realises that this limit $\tau/\kappa \to \infty$ is an IR limit where confinement takes place. Unless at higher loop order the magnetic mass increases quickly enough with temperature in order to act as an efficient IR cutoff, we cannot get around this problem without actually solving confinement. We believe this to be an important consideration for phase transitions which involve non-abelian gauge fields.

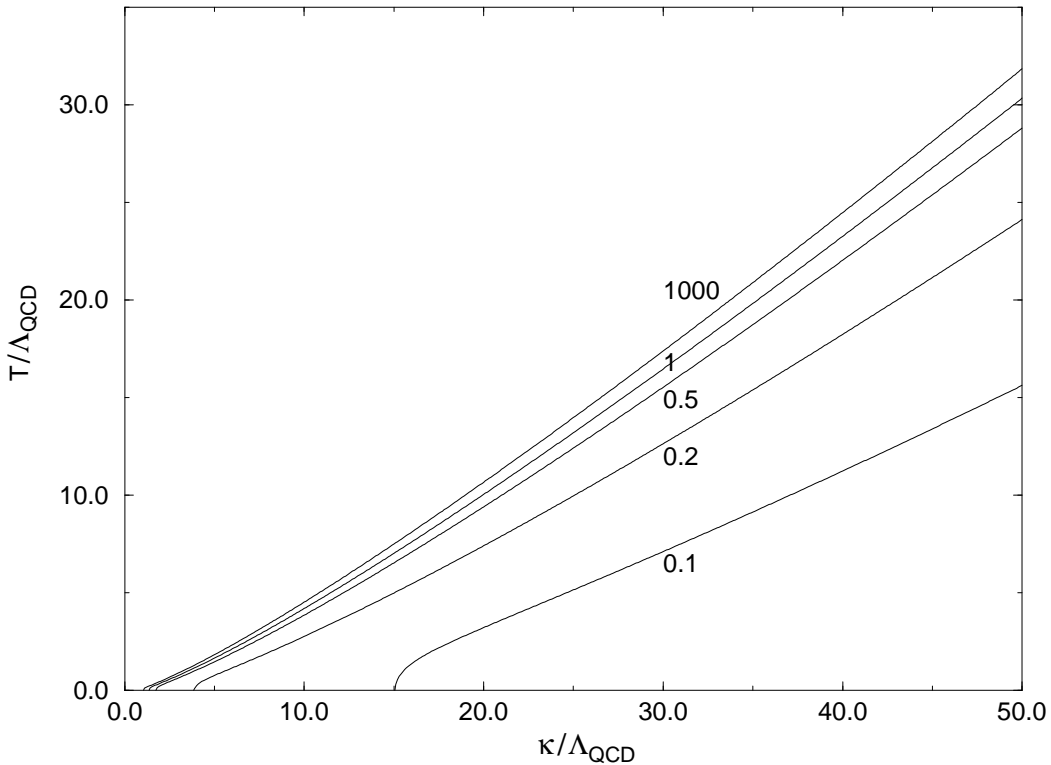

Fig. 2: Contour plot of the running coupling $\alpha_{\kappa,\tau}(\kappa,\tau)$ for QCD with three colours and six fermion flavours. The fermions have been taken massless. Only below the curve $\alpha_{\kappa,\tau} = \infty$ (close to $\alpha_{\kappa,\tau} = 1000$) is the coupling positive and finite.

In the regime $\tau \gg \kappa$ the beta functions behave as in a three-dimensional theory so that we designate this as the region where dimensional reduction occurs. Here it is natural, as for $\lambda\phi^4$, to use a different dimensionless coupling, for instance, $u = \alpha_{\kappa,\tau}\frac{\tau}{\kappa}$. However in this case such a reparametrization cannot remove the pole and will not give a different behaviour.

If we allow the momentum scale to change with temperature simultaneously, the high-temperature limit can be taken in many ways. In the region $\tau \gg \kappa$ the shape of the contours is given by $\tau \sim \kappa \ln \frac{\kappa}{\Lambda_{QCD}}$. This characterizes exactly along which paths in the $(\tau, \kappa)$-plane the coupling increases or decreases. For example at a fixed ratio $\tau/\kappa$ (no matter what this ratio is) we eventually find a coupling that decreases like $1/\ln \kappa$, much in the same way as at zero temperature. This is a natural contour to consider for a weak-coupling regime[15] where one could treat the quark-gluon plasma as a perfect gas, as then the thermal average of the momentum of massless quanta at temperature $T$ is proportional to the temperature. However at low momenta the assumption of weak coupling breaks down. Furthermore, instead of considering quantities at the average momentum it is more appropriate to use thermal averages of the quantities themselves as a weighted integral over all momenta[16]. But once again one runs into problems at the low-momentum end as long as we cannot treat the strong-coupling regime.

## DISCUSSION AND CONCLUSIONS

There have been several approaches to trying to ameliorate the IR problems inherent in a quantitative discussion of the electroweak phase transition. One of the most popular has been associated with the resumming of infinite sets of Feynman diagrams, and in particular ring diagrams. This can be done either in the Higgs sector alone or in the full electroweak model where both scalar loops and gauge loops must be summed. Dimensional reduction is then motivated by considerations of the ring-resummed effective potential. Given that dimensional reduction occurs an effective three dimensional theory has been treated using critical phenomena techniques such as $\varepsilon$-expansions. Dimensional reduction occurs when $\frac{T}{M(T)} \gg 1$ (i.e. in the vicinity of a weakly first order or second order phase transition). In this regime the ring-resummed effective potential is a bad approximation to the true effective potential. Consequently it is not particularly consistent to motivate dimensional reduction by considerations of the ring-resummed effective potential. Naturally a more desireable approach would be to derive whether dimensional reduction occurs or not using a formalism that treats in a perturbatively reliable manner not only the zero temperature regime and the "deep IR" regime (where dimensional reduction occurs) but also the entire crossover between these qualitatively different regimes. Environmentally friendly renormalization, as briefly reviewed in this contribution, is precisely such a formalism.

We have explained how environmentally friendly renormalization gives a complete description of the finite temperature behaviour of $\lambda \phi^4$ theory which is closely related to the Higgs sector of the standard model. We have seen also that there is a significant difference when treating the magnetic sector of non-abelian gauge fields. The implication in this case is that the RG we have employed is not sufficiently environmentally friendly and is manifest in the breakdown in perturbation theory as one runs into the IR. There are several possibilities for making it more environmentally friendly. If the suggestion above that the crossover is to confined degrees of freedom is true then an RG must be implemented that explicitly takes account of this fact, i.e. a reparametrization must be found that captures this crossover. If such a reparametrization could be found then it would provide a completely perturbative description of the crossover between asymptotic freedom and confinement. If in contrast the problem is resolved by the appearance of a magnetic mass in higher orders in perturbation theory then the $\beta$ functions should depend on this mass. The above considerations would apply to the electroweak model in the symmetric phase. In the broken phase the Higgs mechanism gives a mass and therefore an IR cutoff to the gauge bosons. The extent to which this can act as an efficient IR cutoff as the phase transition is approached is not clear without a more detailed analysis. Certainly we believe that environmentally friendly renormalization is an extremely useful tool in the analysis of the electroweak phase transition, and more generally in the context of phase transitions in particle physics and cosmology.

## ACKNOWLEDGEMENTS

DOC would like to thank NWO for financial support during a visit to the University of Amsterdam.